\documentclass[twocolumn,showpacs,preprintnumbers,amsmath,amssymb,floatfix]{revtex4}

\usepackage{graphicx}
\usepackage{dcolumn}
\usepackage{bm}
\usepackage{color}

\newcommand{\be}{\begin{equation}}
\newcommand{\ee}{\end{equation}}
\newcommand{\ba}{\begin{eqnarray}}
\newcommand{\ea}{\end{eqnarray}}

\newcommand{\Ms}{M_{\odot}}
\newcommand{\m}{\langle}
\newcommand{\M}{\rangle}

\def\ltsima{$\; \buildrel < \over \sim \;$}
\def\simlt{\lower.5ex\hbox{\ltsima}}
\def\gtsima{$\; \buildrel > \over \sim \;$}
\def\simgt{\lower.5ex\hbox{\gtsima}}
\begin{document}

\title{Probing white dwarf interiors with LISA: periastron precession in double white dwarfs}

\author{B. Willems$^1$, A. Vecchio$^{1,2}$, V. Kalogera$^1$} 
\affiliation{$^1$Northwestern University, Department of Physics and
  Astronomy, 2145 Sheridan Road, Evanston, IL 60208, USA \\
  $^2$School of Physics and Astronomy, University of Birmingham, 
  Edgbaston, Birmingham B15 2TT, UK }
 
\begin{abstract}
In globular clusters, dynamical interactions give rise to a population of eccentric double white dwarfs detectable by the Laser Interferometer Space Antenna (LISA) up to the Large Magellanic Cloud. In this Letter, we explore the detectability of periastron precession in these systems with LISA. Unlike previous investigations, we consider contributions due to tidal and rotational distortions of the binary components in addition to general relativistic contributions to the periastron precession. At orbital frequencies above a few mHz, we find that tides and stellar rotation dominate, opening up a possibly unique window to the study of the interior and structure of white dwarfs.
\end{abstract}


\maketitle

\emph{Introduction.}-- Binary stars have long been recognized  as unique astrophysical laboratories for the study of physics and cosmology \citep{2006Ap&SS.304....5G}. When one of the binary components is a compact object, measurements of its interactions with the orbital companion provide a wealth of information on the compact object and the properties of matter under extreme conditions. 

The Laser Interferometer Space antenna (LISA; \citep{Bender}) will survey the whole galactic population of  binaries consisting of two white dwarfs (WDs) with periods $\simlt 6$ hr by monitoring their gravitational wave (GW) emission. LISA will individually resolve $\sim 10^4$ double white dwarfs (DWDs) \citep{2001A&A...365..491N} and will be particularly effective in discovering short-period ($\simlt 30$ min) systems -- in stark contrast to current and planned electromagnetic observatories. LISA will therefore allow astrophysical studies of outstanding questions in compact object binary physics, such as dynamically unstable mass transfer, accretion, and type Ia supernova formation \citep{2006AIPC..873..397N}.

So far, investigations of DWDs have focused on the formation and properties of binaries with circular orbits, which dominate the Galactic DWD population. However, \citet{dwd} recently predicted the existence of a sub-population of DWDs consisting of eccentric binaries formed through dynamical interactions in globular clusters. 
Unlike present and planned electromagnetic telescopes, LISA will be able to detect these systems to distances as far as the Large Magellanic Cloud, providing a unique opportunity to study degenerate matter through the imprint of tidal effects in the detected GW signal.  

Tidal interactions in close binaries couple the spins of the component stars to the orbital motion, driving binaries to a state of minimum kinetic energy in which the orbit is circular, the stellar spin angular momenta are aligned with the orbital angular momentum, and the stellar rotation rates are synchronized with the orbital motion. The efficiency of this process depends strongly on the mode of energy dissipation in the stellar interiors, which is reasonably well understood for non-degenerate stars, but an open question for WDs \citep{2004MNRAS.350..113M}. 
However, in eccentric binaries, tides also cause a non-dissipative precession of the periastron of the orbit. This ``apsidal motion" is caused by perturbations in the gravitational field due to the tidal distortions of the binary components. The apsidal motion depends primarily on the internal mass distribution of the stars, and takes place on considerably shorter time scales than dissipative spin-orbit coupling. The shorter time scales facilitate the inference of apsidal-motion rates from electromagnetic and GW observations.      

In addition to tides, rotational distortions of binary stars and general relativity (GR) also contribute to the apsidal motion in eccentric binaries. While the rotational contribution depends on the internal mass distribution in a similar way as the tidal effect, the GR contribution depends only on the total system mass and orbital elements. For non-degenerate stars, comparisons between theoretically predicted apsidal-motion rates and observationally inferred rates have long served as a critical test of theories of stellar structure and evolution \citep{1958ses..book.....S, 1993A&A...277..487C, 2002A&A...388..518C}. 

In this Letter, we examine the physics accessible by measuring the apsidal motion of eccentric DWDs with LISA. General relativistic apsidal motion has already been proposed as a tool to derive the total system mass of eccentric neutron star (NS) binaries \citep{2001PhRvL..87y1101S}. Applications to WD binaries have so far not been considered, nor 
have the imprint on GWs of the periapse precession induced by tides and stellar rotation.

\emph{Apsidal motion.}--- 
We briefly summarize the equations governing the tidal, rotational, and GR contributions to the apsidal motion in eccentric binaries. For this purpose, we consider a binary consisting of two uniformly rotating stars with masses $M_{1,2}$, radii $R_{1,2}$, and rotational angular velocities $\Omega_{1,2}$.The stellar rotation axes are assumed to be perpendicular to the orbital plane \footnote{ Generalizations of the equations governing the periastrion precession allowing for inclined rotation axes have been considered by e.g. \citet{1985SvAL...11..224S}.}. We furthermore let $P$ be the orbital period, $a$ the semi-major axis, $e$ the orbital eccentricity, and $\gamma$ the argument of the periastron. 

The contribution to the apsidal motion from the tidal distortion of the binary components is most commonly determined under the assumption that the orbital and rotational periods are long compared to the periods of the free oscillation modes of the component stars \citep{1938MNRAS..98..734C, 1939MNRAS..99..451S, 2001A&A...373..173S}. Under this assumption, the rate of secular apsidal motion due to the dominant quadrupole tides raised in star $i$ ($i=1,2$) is given by
\begin{equation}
\dot{\gamma}_{\rm tid,i} = {{30\,\pi} \over P} 
  \left( {R_i \over a} \right)^5 {M_{3-i} \over M_i}\, 
  {{1 + {3\over 2}\, e^2 + {1\over 8}\, e^4}
  \over {\left( 1-e^2 \right)^5}}\,k_i, \label{tid}
\end{equation}
where $k_i$ is the quadrupolar apsidal-motion constant of star $i$. When the orbital and rotational periods are of the order of the free oscillation modes of the binary components, deviations from Eq.~(\ref{tid}) arise due to the increasing role of stellar compressibility on the tidal displacement field for higher tidal forcing frequencies and due to the occurrance of resonances between dynamic tides and nonradial stellar oscillations \citep{2001A&A...373..173S}.

The apsidal-motion constants $k_i$ measure the degree to which mass is concentrated towards  the stellar center and are determined by numerical integration of the equation of Clairaut (for details see, e.g., \citep{1939MNRAS..99..451S, 2002A&A...382.1009W}). The constants are unaffected by dissipative effects as long as the tidal forcing is not in resonance with any of the WDs' nonradial stellar oscillation modes \citep{2001A&A...373..173S, 2003A&A...397..973W}. In the limiting cases where the stars are approximated by point masses or equilibrium spheres with uniform mass density, the constants $k_i$ take the values 0 and 0.75, respectively. For more realistic stellar models, the constants take values between these two extremes.

Rotation contributes to the apsidal motion through the rotational quadrupole distortion caused by the centrifugal force \citep{1939MNRAS..99..451S}. The corresponding rate of secular apsidal motion depends on $R_{1,2}$ and $k_{1,2}$ in a similar way as the tidal contribution to the apsidal-motion rate, but has a different dependence on $M_{1,2}$ and $e$:
\begin{equation}
\dot{\gamma}_{\rm rot,i} = {{2\,\pi} \over P}
  \left( {R_i \over a} \right)^5 {{M_1 + M_2} \over M_i}\, 
  {{\left( {\Omega_i/\Omega} \right)^2} \over 
  {\left( 1-e^2 \right)^2}}\, k_i, \label{rot}
\end{equation}
where $\Omega = 2\,\pi/P$ is the mean motion. 

The GR contribution to the apsidal-motion rate differs from the tidal and rotational contributions in that it is independent of the radii and internal structure of the binary components. At the leading quadrupole order, the GR apsidal-motion rate is given by 
\begin{equation}
\dot{\gamma}_{\rm GR} = {{2\,\pi} \over P}\, 
  {{3\,G} \over c^2}\, {{M_1+M_2} \over 
  {a \left( 1-e^2 \right)}}, \label{gr}
\end{equation}
where $G$ is the Newtonian gravitational constant, and $c$ the speed of light \citep{Levi}. 

\begin{figure}
\begin{center}
\resizebox{8.0cm}{!}{\includegraphics{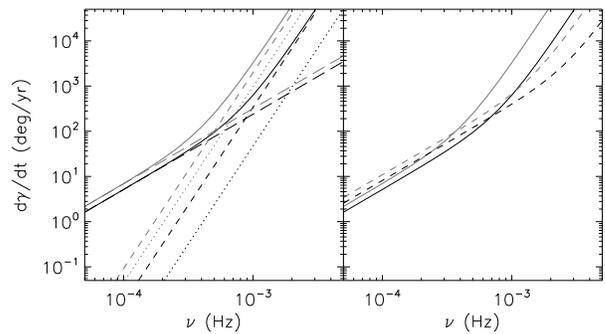}}
\caption{{\it Left:} Tidal (short-dashed), rotational (dotted), GR (long-dashed), and total (solid) apsidal-motion rate for DWDs with $M_1=M_2=0.3\,M_\odot$. {\it Right:} Total apsidal-motion rate for DWDs with $M_1=M_2=0.3\,M_\odot$ (solid) and $M_1=M_2=0.6\,M_\odot$ (short-dashed). Both panels show results for $e=0.01$ (black) and $e=0.5$ (grey), and $k_1=k_2=0.1$.
}
\label{f:apsrate}
\end{center}
\end{figure}

In Fig.~\ref{f:apsrate}, the tidal, rotational, GR, and total apsidal-motion rate are shown as functions of the orbital frequency $\nu=1/P$ for different orbital eccentricities and conservative WD component masses of $0.3$ and $0.6\,M_\odot$. 
The tidal and rotational contributions are determined using Nauenberg's \citep{1972ApJ...175..417N} zero-temperature mass-radius relation, and setting $k_1 = k_2 = 0.1$. These $k_1$ and $k_2$ are appropriate for cool WD models of $0.3\,M_\odot$ and $0.6\,M_\odot$; the dependence of $k_1$ and $k_2$ on the WD mass and temperature will be explored in more detail in a separate investigation.
The WD rotational angular velocities are furthermore assumed to be synchronized with the orbital angular velocity at periastron.  
Since the tidal effects usually dominate the rotational effects, this assumption does not affect the main conclusions of the calculation.
 
It is evident that the total apsidal-motion rate is substantial throughout the entire LISA band, even for eccentricities as low as $e \simeq 0.01$: at $\nu \simeq 0.5$ mHz, periastron precession already induces a phase shift 
in the GW signal of more than $2\pi$ over an observation time $T_\mathrm{obs} = 5$ yr (the current minimum mission lifetime requirement) and therefore becomes potentially detectable. The phase shift is even larger at higher orbital frequencies. 
Equally striking is the dominance of tides and stellar rotation at frequencies above a few mHz. \citet{dwd} have shown eccentric DWDs in this frequency range to be detectable by LISA up to distances as far as the Large Magellanic Cloud, opening up 
new avenues for GW astrophysics of DWDs. The tidal and rotational contributions to $\dot{\gamma}$ furthermore decrease with increasing mass of the WDs due to the smaller radii of more massive WDs. 
At low frequencies, where GR effects dominate, the apsidal-motion rate increases with increasing mass of the WDs.

\emph{LISA observations.}--- Periastron precession leaves a signature in the GW forms of eccentric DWDs by modifying the phase of the signal recorded by laser interferometers. LISA can therefore probe into the structure of DWDs by observing the apsidal motion due to the above contributions. Here, we first show conceptually how one can measure $\gamma$ and the parameters that drive its evolution; next we explore how accurately these parameters can be measured. Throughout this discussion we model GW radiation at the leading Newtonian quadrupole order 
(post-Newtonian corrections are negligible in this frequency and mass range), 
and model LISA following~\citep{2004PhRvD..69h2005B}.

The signal from an eccentric DWD in the LISA frame can be schematically written as 
\be
h(t)  = \frac{\sqrt{3}}{2}\,\sum_n \left[F_+(t) h^{+}_n(t) + F_\times(t) h^{\times}_n(t)\right],
\label{e:h}
\ee
where $F_{+,\times}$ are the antenna beam patterns (that depend on the source right ascension $\alpha$ and declination $\delta$, and the wave polarization $\psi$ at a constant reference time), and  
\ba
h^{+}_n(t) & = & A \Bigl\{-(1 + \cos^2\iota)u_n(e) \cos[n\phi(t) + 2 \gamma(t)] 
\nonumber \\ 
& & -(1 + \cos^2\iota) v_n(e) \cos[n\phi(t) - 2 \gamma(t)]
\nonumber \\
 & & + \sin^2\iota\, w_n(e) \cos[n\phi(t)] \Bigr\},
\label{e:h+}\\
h^{\times}_{n}(t) & = & 2 A \cos\iota \Bigl\{u_n(e) \sin[n\phi(t) + 2 \gamma(t)] 
\nonumber\\
& & + v_n(e) \sin([n\phi(t) - 2 \gamma(t)]) \Bigr\}\,.
\label{e:hx}
\ea
Here, $\phi(t)$ is the Doppler shifted orbital phase $\phi_\mathrm{orb} = 2\pi \nu t + \pi \dot{\nu} t^2 + {\cal O}(t^3) + \phi_0$, where $\dot{\nu}$ is the orbital frequency derivative and $\phi_0$ an arbitrary initial phase; $\iota$ is the constant source inclination angle, $A = (2 \pi \nu)^{2/3} {\cal M}^{5/3}/d$ the GW amplitude, ${\cal M}$ the chirp mass, and $d$ the distance to the source; $u_n(e)$,  $v_n(e)$, and $w_n(e)$ are linear combinations of the Bessel functions of the first kind $J_{n}(ne)$, $J_{n\pm 1}(ne)$ and $J_{n\pm 2}(ne)$; explicit expressions can be derived using Eqs.~(7) and~(10) in \citep{2004PhRvD..69h2005B}. 

In the absence of periastron precession, radiation is emitted at multiples $n$ of the orbital frequency $\nu$, but when periastron precession is present, each of these harmonics is split into a triplet with frequencies $n\nu \pm \dot{\gamma}/\pi$ and $n\nu$, and amplitudes $u_n(e)$, $v_n(e)$ and $w_n(e)$, respectively.
As already noted by~\citep{2001PhRvL..87y1101S}, the observation of any two ``emission lines'', allows us to derive the orbital {\em and} the apsidal-motion frequency. In practice, for typical DWD eccentricities 
($e \simlt 0.5$; see~\citep{dwd}), $|u_n(e)| \gg |v_n(e)|\,,|w_n(e)|$, so that LISA will primarily rely on observations of GWs at frequencies $n \nu +  \dot{\gamma}/\pi$ for at least two values of $n$.

We can compute whether and how accurately periastron advance can be measured by computing the Fisher information matrix (see {\em e.g.}~\cite{2004PhRvD..69h2005B}) associated with the measurement. The signal depends on $\alpha$, $\delta$, $\iota$, $\psi$, $A$, $e$, $\nu$, $\dot{\nu}$, $\dot{\gamma}$, $\phi_0$ and $\gamma_0$ (the argument of periastron at an arbitrary reference time). 
Due to the fact that $\alpha$, $\delta$, $\iota$ and $\psi$ are only weakly correlated with the remaining parameters for an observation lasting several years~\citep{2002ApJ...575.1030T}, we do not include them in our analysis and compute the angle-averaged Fisher information matrix for $T_\mathrm{obs} = 5$ yr. We conservatively consider the first 10 harmonics in Eq.~(\ref{e:h}) and normalize the results to signals detected at an optimal signal-to-noise ratio (SNR) of 10 (the mean-square errors scale as $\approx 1/$SNR). The results are summarized in Table~\ref{t:errors} for the parameters relevant to this investigation.

For $\nu \simlt 1$\,mHz, the apsidal motion becomes progressively harder to measure due to the smaller and smaller phase shift (cf.\ Fig.~\ref{f:apsrate}). At $\nu \approx 0.1$ mHz the estimated error on 
$\dot{\gamma}/\pi$ is greater than $\dot{\gamma}/\pi$ itself and apsidal motion becomes undetectable (the details of course depend on the actual values of $e$, $\nu$, $M_1$, $M_2$ and SNR for the source at hand) 
\footnote{For a DWD with $M_1 = M_2 = 0.6\,\Ms$, the value of ${\dot \gamma}/\pi$ is $\simeq 61\, (1.3)/(1 - e^2)$ nHz for $\nu = 1\, (0.1)$ mHz.}. However, for $\nu \simgt 1$\,mHz, LISA can detect periastron precession and measure ${\dot \gamma}/\pi$ with a relative error of $\sim 1\% - 10\%$, depending on the mass and eccentricity of the binary. This result holds even for small eccentricities $e \sim 0.01$~\footnote{For a binary of two $1.4\,\Ms$ NSs with $e = 0.1$ and $\nu = 1$ mHz, and for $T_{\rm obs}=5$ yr and SNR$=10$, ${\dot \gamma}/\pi \simeq 107$ nHz and $\m({\dot \gamma}/\pi)^2 \M^{1/2} \simeq 1.8$ nHz. Our conclusions on LISA's ability to measure $\dot{\gamma}/\pi$ are therefore more optimistic than those reported by Seto in~\cite{2001PhRvL..87y1101S}. Seto's analysis in fact is based on the ultra-conservative assumption that the apsidal motion must produce a frequency shift by a full bin in order to be observable instead of the more rigorous approach adopted in this paper. The evaluation of ${\dot \gamma}/\pi$ given in Eq. (6) of~\cite{2001PhRvL..87y1101S} furthermore contains a numerical error that underestimates the effect by a factor $\approx 5$.}. 
At these frequencies most systems are also expected to exhibit a detectable change of the orbital frequency.  Assuming that only general relativity affects $\dot{\gamma}$ and $\dot{\nu}$, the combined measurement of $\dot{\gamma}$ and $\dot{\nu}$ allows us to determine the total system mass $M$ and chirp mass ${\cal M}$, as shown in Table~\ref{t:errors2}, and therefore the individual WD masses $M_1$ and $M_2$.

\begin{table}
\begin{center}
\begin{tabular}{lcccccc}
\hline
\hline
						&	$\nu$			& \multicolumn{5}{c}{e}											\\														&	(mHz)			&	0.01		&	0.1		&	0.3		&	0.5		&	0.7		\\
\hline	
						&	0.1				&	0.03		&	0.03		&	0.03		&	0.02		&	0.03		\\
$\m(\Delta e)^2\M^{1/2}$				&	1				&	0.02		&	0.02		&	0.03		&	0.03		&	0.03		\\
						&	3				&	0.05		&	0.05		&	0.04		&	0.04		&	0.03		\\
\hline
						&	0.1				&	17.39	&	1.95		&	1.08		&	0.83		&	1.70		\\
$\m(\Delta {\dot \gamma}/\pi)^2\M^{1/2}/$nHz	&	1		&	14.40	&	1.83		&	1.28		&	1.07		&	0.89		\\
						&	3				&	32.30	&	3.28		&	0.12		&	0.86		&	0.69		\\
\hline
$\m(\Delta \dot{\nu})^2\M^{1/2}/$nHz$^2$	&	1			&	8.56		&	7.62		&	5.27		&	3.64		&	2.12		\\
						&	3				&	8.57		&	8.36		&	6.89		&	4.82		&	2.82		\\
\hline
\end{tabular}
\end{center}
\caption{Measurement accuracy of $e$, ${\dot \gamma}/\pi$, and $ \dot{\nu}$ for selected values of $e$ and $\nu$. The errors are for $T_{\rm obs} = 5$\,yr  and for a source detected with optimal SNR $= 10$.  No assumption is made on the physical mechanism driving the apsidal motion $\dot{\gamma}$ and/or frequency drift $\dot{\nu}$. The error on $\dot{\nu}$ is reported only for systems with $\nu \ge 1$ mHz, where the GR contribution to $\dot{\nu}$ becomes observable.
}
\label{t:errors}
\end{table}
\begin{table}
\begin{center}
\begin{tabular}{lcccccc}
\hline
\hline
						&	$\nu$		& \multicolumn{5}{c}{e}											\\															&	(mHz)		&	0.01		&	0.1		&	0.3		&	0.5		&	0.7		\\
\hline	
						&	0.1			&	19.86	&	2.20		&	1.12		&	0.72		&	0.99		\\
$\m(\Delta M)^2\M^{1/2}/M$			&	1			&	0.35		&	0.05		&	0.04		&	0.07		&	0.11		\\
						&	3 			&	0.13		&	0.02		&	0.04		&	0.07		&	0.11		\\
\hline
$\m(\Delta {\cal M})^2\M^{1/2}/{\cal M}$	&	1			&	0.52		&	0.43		&	0.20		&	0.30		&	1.64		\\
						&	3 			&	0.01		&	0.04		&	0.12		&	0.32		&	1.62		\\
\hline
\end{tabular}
\end{center}
\caption{Relative error on total system mass and chirp mass measurements {\em assuming} only GR contributes to $\dot{\gamma}$ and $\dot{\nu}$, for a 5-yr observation of a $M_1\!=\!M_2\!=\!0.6 M_\odot$ DWD at SNR\,=\,10. 
}
\label{t:errors2}
\end{table}

\emph{Astrophysical implications.}--- 
We have investigated the impact of tides, rotation, and GR on the 
apsidal motion of eccentric DWDs and its signature on the emitted GWs. Based on our present understanding of the astrophysical scenarios~\cite{dwd}, we conclude that LISA will be able to observe the periastron advance for the vast majority of such sources detected. These observations provide a new and unique probe into the internal structure of WDs. 

Tides and stellar rotation strongly dominate the apsidal-motion rate at orbital frequencies above $\approx 1$ mHz, inducing phase shifts much larger than those estimated using only the GR contribution. In GW searches for eccentric binaries, it is therefore essential to include $\dot{\gamma}$ in the signal templates as a phenomenological parameter {\em not bound} by $\dot{\gamma}_\mathrm{GR}$ in order to not bias LISA surveys against eccentric DWDs. In the interpretation of the data, neglecting tidal and rotational contributions would lead to an overestimate of the total system mass derived from $\dot{\gamma}$. This will likely induce a misclassification of GW sources as NS rather than DWD binaries, thus affecting the ratios of different populations of compact object binaries that hold essential signatures of stellar evolution and binary formation mechanisms. On the other hand, the dependence of $\dot{\gamma}$ on $R_{1,2}$, $M_{1,2}$, and $k_{1,2}$ provides a new window into the internal structure of WDs and the astrophysics of such stars, but poses a severe degeneracy problem for the extraction of astrophysical information from measured apsidal-motion rates. More refined theoretical modeling is therefore needed in preparation of the LISA mission to fully characterize the dependence of $\dot{\gamma}$ on the WD physical parameters and to identify routes to untangle them. Even though we have here focused on DWDs, our results also apply to WDs with NS companions. These sources are in fact much ``cleaner'' probes of WD physics since the tidal and rotational distortions of the NS contribute negligibly to the apsidal motion. In this case, the apsidal-motion rate therefore carries the unique signature of only one WD rather than two.

At frequencies $\simlt 1$\,mHz, GR effects dominate the precession rate and measurements of $\dot{\gamma}$ allow the determination of the total system mass \citep{2001PhRvL..87y1101S}.  For $\nu \simgt 0.5$\,mHz \citep{dwd}, the radiation reaction may also cause a measurable drift in the orbital frequency. Assuming that there is no significant contribution from tidal and/or magnetic spin-orbit coupling, measuring $\dot{\nu}$ with LISA yields the source distance as well as the chirp mass  \citep[e.g.][]{Stroeer05}. In the (small) regime where general relativity dominates the apsidal motion and $\dot{\nu}$ is detectable, the combined knowledge of the chirp mass and total system mass yields the masses of the individual WD components. For eccentric NS-NS binaries -- which have negligible tidal/rotational distortions and tidal/magnetic spin-orbit coupling  throughout the LISA band 
and a stronger $\dot\nu$ than WD binaries -- the measurement of the individual masses at $\approx 10\%$ level should be routine.

\emph{Acknowledgments.}--- This work is partially supported by a Packard Foundation Fellowship, a NASA BEFS grant (NNG06GH87G), and a NSF CAREER grant (AST-0449558) to VK. We are grateful to Brad Hansen and Chris Deloye for providing theoretical WD models.

\end{document}